\documentclass[acmsmall,screen]{acmart}

\AtBeginDocument{%
  }

\setcopyright{none}
\acmYear{2026}
\renewcommand\footnotetextcopyrightpermission[1]{}

\usepackage{tabularx}
\usepackage{tcolorbox}
\tcbuselibrary{skins}
\usepackage{makecell}
\usepackage{threeparttable}
\usepackage{multirow}
\usepackage{setspace}
\usepackage{cleveref}
\usepackage{subcaption}
\usepackage{booktabs}
\usepackage{listings}

\usepackage{enumitem}
\setlist{topsep=3pt, itemsep=1pt, parsep=0pt, partopsep=0pt}

\lstdefinestyle{promptstyle}{
  basicstyle=\footnotesize\ttfamily,
  breaklines=true,
  columns=fullflexible,
  keepspaces=true,
  frame=single,
  framerule=0.4pt,
  rulecolor=\color{black},
  framesep=4pt,
  xleftmargin=1em,
  xrightmargin=1em,
  aboveskip=0.6em,
  belowskip=0.6em,
}

\newtcolorbox{findingbox}{
  colback=white, colframe=black, boxrule=0.8pt,
  arc=0pt, left=6pt, right=6pt, top=4pt, bottom=4pt,
  before skip=2pt, after skip=2pt
}

\definecolor{champblue}{HTML}{2A78D6}
\definecolor{champorange}{HTML}{EB6834}
\newtcolorbox{codecard}[2][champorange!20]{
  enhanced,
  colback=white, colframe=black!75,
  boxrule=0.5pt, arc=1pt,
  left=4pt, right=4pt, top=5pt, bottom=4pt,
  attach boxed title to top left={yshift=-2pt, xshift=6pt},
  boxed title style={colback=#1, boxrule=0pt, arc=1pt,
    left=3pt, right=3pt, top=1.5pt, bottom=1.5pt},
  coltitle=black,
  title={\makebox[90pt][l]{\scriptsize\bfseries #2}},
}

\begin{document}

\title{Not All Correct Fixes Run Alike: An Empirical Study of Patch
  Performance in Repository-Level Issue Resolution}

\author{Han Peng}
\affiliation{%
  \institution{The State Key Laboratory of Blockchain and Data Security, Zhejiang University}
  \city{Hangzhou}
  \country{China}
  }
\affiliation{%
  \institution{Hangzhou High-Tech Zone (Binjiang) Institute of Blockchain and Data Security}
  \city{Hangzhou}
  \country{China}
  }
\email{awdhanpeng@zju.edu.cn}

\author{Xing Hu}
\affiliation{%
  \institution{The State Key Laboratory of Blockchain and Data Security, Zhejiang University}
  \city{Hangzhou}
  \country{China}
  }
\email{xinghu@zju.edu.cn}

\author{Feifei Niu}
\affiliation{%
  \institution{Graz University of Technology}
  \city{Graz}
  \country{Austria}
  }
\email{feifeiniu96@gmail.com}

\author{Huan Li}
\affiliation{%
  \institution{The State Key Laboratory of Blockchain and Data Security, Zhejiang University}
  \city{Hangzhou}
  \country{China}
  }
\email{huanlee@zju.edu.cn}

\author{Xin Xia}
\correspondingauthor
\affiliation{%
  \institution{The State Key Laboratory of Blockchain and Data Security, Zhejiang University}
  \city{Hangzhou}
  \country{China}
  }
\email{xin.xia@acm.org}

\renewcommand{\shortauthors}{Han Peng et al.}

\begin{abstract}
Repository-level issue-resolution benchmarks such as SWE-bench evaluate submitted patches primarily based on whether they pass the correctness tests, leaving the runtime performance of different test-passing patches largely unexplored. Evaluating patch performance is essential but challenging since designing performance workloads for each benchmark instance is costly. However, these benchmarks already provide a natural candidate: the built-in \emph{FAIL\_TO\_PASS} (F2P) tests, which exercise the behavior affected by the issue fix. We therefore investigate whether the provided F2P tests can be repurposed to evaluate patch performance. Overall, we choose four frontier models, Claude Opus~5, GPT~5.6, DeepSeek V4 Pro, and Qwen~3.8 Max, paired with SWE-agent to resolve the pre-filtered measurable corpus from verified splits of SWE-bench and SWE-bench-Live, yielding 262 cases with at least two different correct solutions, including the developer ground-truth patch. We then measure each solution's runtime performance in instruction count and wall-clock time on the native F2P test, and use Welch ANOVA instead of the RSD proposed in prior work as a more robust statistical discrimination criterion. We find that this methodology exhibits notable discrimination capability in runtime performance and identifies performance differences among patches in almost half of the instances. Our empirical study shows that the runtime performance of generated patches is independent of the issue resolution rate of LLMs, with nearly reversed model rankings. The case study also shows that the performance analysis reveals architectural regression in some patches that the correctness evaluation overlooked. These findings suggest that resolution rate does not adequately characterize LLMs' capability for repository-level issue resolution, motivating future evaluations to consider patch performance alongside it.
\end{abstract}



\maketitle

\section{Introduction}
\label{sec:intro}

Repository-level issue-resolution benchmarks such as SWE-bench~\cite{jimenez2024swe} and SWE-bench-Live~\cite{zhang2026swe} have become widely used to evaluate the capability of large language models (LLMs) to resolve real-world software issues. For each instance, a solver receives the problem statement, works in the repository at the base commit, and generates a patch. The benchmark then checks whether the applied patch passes the test suites, especially the \emph{FAIL\_TO\_PASS} (F2P) tests, which fail on the unfixed repository and pass after the repair, to determine whether the issue is resolved. This evaluation, however, focuses only on functional correctness. Passing the tests does not indicate how efficiently the patch executes: different correct patches may satisfy the same tests while exhibiting substantially different performance costs. Runtime performance is therefore an important dimension of patch quality that current benchmarks fail to capture.

Recently, code performance has drawn increasing attention alongside correctness. However, the research community has not examined the runtime behavior of generated patches for repository issues related to functional correctness. One line of related work evaluates the runtime performance of code generated from functional requirements, primarily at the function or single-file level~\cite{huang2024effibench,peng2025coffe,li2024assessing}. In contrast, our study concerns issue resolution in real-world repositories. The other related line focuses on repository-level code performance optimization~\cite{he2025sweperf,shetty2026gso,ma2026swefficiency}. These studies usually carry performance workloads to measure speedup and aim solely to improve runtime performance rather than repair existing functional issues. Overall, our study differs from both lines of research: we focus on whether LLMs produce genuinely different correct patches for the same repository-level functional issues, particularly in runtime performance.

Runtime performance analysis requires workloads to measure the patched code. The repository-level issue-resolution tasks packaged by benchmarks such as SWE-bench, however, provide only test suites for correctness, with none for performance analysis. Nevertheless, we note that the built-in F2P test exercises the exact behavior the fix changes, so running it repeatedly and recording its execution cost seems to provide a free performance evaluation. We therefore investigate whether this F2P test can evaluate the runtime performance of different correct patches and effectively reveal performance differences among them. We build our research corpus from the verified splits of SWE-bench and SWE-bench-Live, through a pre-filtering process that removes the instances unsuitable for performance measurement and retains 339 cases. After that, we employ four frontier LLMs, namely \textbf{Claude Opus~5}, \textbf{GPT~5.6}, \textbf{DeepSeek V4 Pro}, and \textbf{Qwen~3.8 Max}, each paired with the widely used SWE-agent framework~\cite{yang2024swe} to resolve the retained instances. We then collect the correct patches produced by these models, analyze the diversity and similarity landscape among them and the developer ground-truth patches, and repeatedly run the F2P test on each to record instruction counts and wall-clock times for performance analysis. As the statistical discrimination criterion, we employ the Welch analysis of variance (ANOVA)~\cite{welch1951comparison} in place of the Relative Standard Deviation (RSD) proposed by prior work~\cite{peng2025coffe}, considering that the latter ignores the within-solution variance and is therefore not robust enough, especially when the number of solutions under comparison varies across cases, or when several solutions form closely related subgroups.

Our empirical study yields the following findings. \textbf{First}, in the filtered corpus, 77.29\% of instances have multiple different correct patches under static comparison, indicating the widespread solution diversity across LLMs and the ground truth. \textbf{Second}, the F2P tests provide meaningful discrimination signals for runtime performance: in 41.98\% and 34.73\% of cases, different solutions exhibit statistically significant performance differences in instruction count and wall-clock time, respectively, suggesting that the F2P tests could serve as a proper starting point for performance analysis. \textbf{Third}, the discrimination criteria of Welch ANOVA and RSD show inconsistent detection behavior, and the latter is likely to overlook cases with genuine runtime differences while retaining cases without meaningful distinctions. \textbf{Fourth}, the resolution rate of issues and the runtime performance of the generated patches are essentially uncorrelated for the compared LLMs, with nearly reversed model rankings:
\begin{itemize}
  \item For the issue resolution rates, Claude (87.32\%) $\gg$ Qwen (83.19\%) $>$ GPT (82.01\%) $\approx$ DeepSeek (81.42\%);
  \item For the speedup against the ground truth, GPT (97.53\%) $\approx$ DeepSeek (97.89\%) $>$ Claude (98.42\%) $\gg$ Qwen (101.68\%), with the smaller the better;
  \item For the rate of producing the fastest solution, DeepSeek (26.3\%) $>$ GPT (24.7\%) $\gg$ Qwen (15.3\%) $>$ Claude (15.0\%), against 18.7\% of the ground truth.
\end{itemize}
The model with the highest resolution rate produces only the third-most efficient patches, and the developer ground-truth patches show no systematic performance advantage over model-generated patches. \textbf{Lastly}, the case study shows that performance comparisons reveal architectural regressions that correctness evaluation cannot detect: three repair strategies vary performance by more than twice, and the two costly ones silently reintroduce the heavy dependency the developer deliberately removed. Based on these findings, we call for future research to focus not only on issue-resolution rates but also on the performance of generated patches for real-world repository issues.

This paper makes the following contributions:
\begin{itemize}
  \item \textbf{An empirical study of runtime performance for LLM-generated patches.} We conduct a large-scale empirical study of correct patches generated by LLMs and by developers. We investigate whether LLMs produce essentially different correct patches, particularly in runtime performance, for repository-level functional issue resolution, and we answer this question on two SWE-bench-family benchmarks with four frontier LLMs using the established agent framework.
  \item \textbf{A practical approach to measure patch performance.} We propose a measurement methodology that repurposes the benchmarks' built-in F2P tests as performance workloads, revealing statistically meaningful performance differences among correct solutions in nearly half of the studied cases. 
  \item \textbf{A principled criterion to distinguish performance differences.} To our knowledge, we are the first to apply Welch ANOVA to compare the runtime performance of LLM-generated patches for repository issues. It accounts for both between-solution and within-solution variation, effectively revealing genuine runtime differences among correct patches.
  \item \textbf{Findings and implications for patch performance evaluation.} Our empirical study shows that correct patches vary widely across LLMs and developers. It also shows that not only the resolution rate but also the runtime performance of generated patches reflects LLMs' ability to resolve real-world repository issues, with their model rankings nearly reversed, highlighting the need to account for patch runtime performance.
\end{itemize}


\section{Preliminaries: From F2P Test to a Measurable Corpus}
\label{sec:design}

This section establishes the basis for our performance evaluation. We introduce the datasets and task (\S\ref{sec:setting}), filter the raw datasets into a corpus amenable to performance evaluation (\S\ref{sec:filtering}), and use LLM agents to solve the corpus and merge their patches into separate solutions (\S\ref{sec:solving}).

\subsection{Datasets and Task}
\label{sec:setting}

In this study, we analyze patch performance in repository-level issue-resolution benchmarks. The most widely used benchmark is SWE-bench~\cite{jimenez2024swe}, which harvests real issues and their resolving pull requests from popular open-source Python repositories. SWE-bench-Live~\cite{zhang2026swe} updates continuously with more recent pull requests created after the knowledge cutoffs of contemporary models, mitigating data contamination. We conduct our study using the verified splits of both benchmarks.

Each instance in these benchmarks corresponds to a real-world repository-level issue. It includes a problem statement describing the reported issue, a base commit specifying the repository state before the fix, the developer's gold patch, and a test patch containing the F2P tests that encode the expected behavior, along with a \emph{PASS\_TO\_PASS} (P2P) regression test suite. Each instance also includes a pre-built Docker execution environment in which the submitted patch is applied, and the tests are executed. To address an instance, a solver reads the problem statement, operates in the repository at the base commit, and submits a diff patch; the official evaluation infrastructure applies the patch and runs the test suite. An instance is considered resolved if all F2P tests pass while the P2P regression tests remain passing. 

The central correctness criterion is the F2P test itself. It fails on the unfixed repository and passes only after the issue is repaired. The task is therefore test-driven; correctness is mainly determined by whether the submitted patch passes the F2P test, rather than by whether it reproduces the developer's reference patch. The evaluation criterion only requires that the issue be fixed, not the specific implementation used to achieve it. So multiple genuinely different patches may pass the same F2P tests. These alternative fixes can exhibit measurably different performance once applied to the repository. This performance dimension is invisible to the correctness protocol, yet real and varying across admissible fixes, and the benchmark provides no direct way to observe it.

An intuitive way to measure performance, however, appears readily available. The F2P test exercises exactly the behavior affected by the fix changes and therefore serves as an approximate stress workload. By repeatedly executing the test under instrumentation, we can collect performance metrics during execution and evaluate each patch on each benchmark instance without requiring additional artifacts. Note that the benchmark also provides the P2P suite to guard against correctness regressions; however, we do not extend performance measurement to it. The P2P tests typically far outnumber the F2P ones, and most do not exercise the issue-fixed code path. Including them would multiply the cost while contributing little to the performance analysis.

We acknowledge that an F2P test typically combines data preparation, invocation of the tested behavior, and assertion in a single body, so measuring the whole test captures all three phases, while only the invocation is patch-sensitive. However, in the ideal case, data preparation and assertion cost the same for every solution. Thus, including them shifts every solution's cost by the same amount, diluting the measured differences without biasing the comparison. In reality, the execution introduces run-to-run noise, while the measurement protocol averages it out through repeated executions, so it favors no solution. As a result, the ordering among solutions still remains trustworthy when the difference is statistically significant. The residual effect of noise is confined only to solutions whose costs are close, where a raw reading cannot tell whether the gap is real.

\subsection{Filtering for Measurability}
\label{sec:filtering}

In practice, however, not every instance supports this straightforward usage. During corpus construction, we therefore impose three constraints on the instances. First, the F2P test must constitute a single unit of measurement, so that measurement yields one comparable reading per instance. Second, it must sustain repeated execution within one process to suppress run-to-run noise. Third, it must be free of input and output operations, whose variance would drown out the CPU-level signals the measurement relies on. We filter the datasets against these constraints, using one static rule for the first and dynamic probing for the latter two, and retain 339 instances as our corpus.

\subsubsection{Static Constraint: One Test per Instance}

We restrict our scope to instances whose F2P field contains a single test. When an instance lists multiple F2P tests, each test would yield its own measurement; roughly merging them into a single reading, however, is inappropriate for analysis, because it hides each test's contribution and weakens the ability to expose performance differences between solutions. For simplicity, we therefore leave these instances out of scope and use the size of the F2P list as the criterion, excluding instances with more than one F2P test.

\subsubsection{Dynamic Probing: Repeatability and I/O Freedom}
\label{sec:probing}

The remaining two constraints concern how the test behaves during execution, and no field in the dataset can answer them. We establish both constraints by dynamically probing each instance.

\begin{figure}[t]
  \centering
  \includegraphics[width=\linewidth]{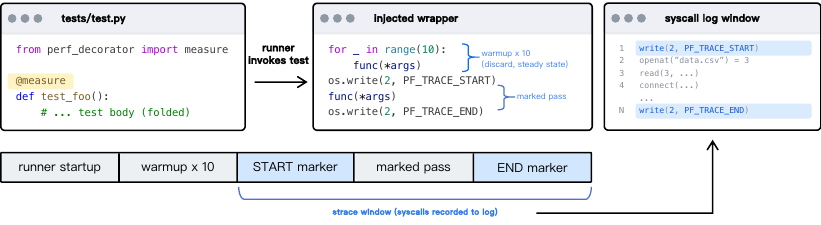}
  \caption{The decorator-based probing, simplified from the injected wrapper.}
  \label{fig:probing}
\end{figure}

\begin{itemize}
  \item \emph{In-process repeatability.} Conventional performance evaluation organizes execution at two levels: the inner loop repeats the measured function to accumulate signal, lifting the measurement quantity above the noise floor, and the outer loop repeats that process as independent replicates to expose run-to-run variation~\cite{georges2007statistically}. In-process repetition is essential, since a single execution of the F2P test is typically very fast and cannot yield an accurate measurement on its own. The test must therefore support repeated execution within a single process; a test that breaks under repetition cannot be measured, and is excluded.
  \item \emph{I/O-free execution.} Some F2P tests involve input/output operations (such as network and disk access) which would drown out the performance signals. Note that we execute the F2P test as a complete unit, since separating its invocation of the target function from both data preparation and assertion is nontrivial and beyond our research scope. As a result, we exclude F2P tests that perform I/O anywhere in their bodies.
\end{itemize}
Both rules demand executing the F2P test and analyzing its behavior. 
The benchmark provides each instance's native test command and its test identifiers. Appending the F2P identifier to the native command runs exactly that test as a single point, so probing reuses the official execution infrastructure throughout. On top of it, we devise a decorator-based probing method that exploits a Python language feature. It requires no modification to the test's own logic, keeping our method generic across the benchmark's diverse repositories (shown in Fig.~\ref{fig:probing}). 

Specifically, for each instance, we first locate its F2P function in the test file that includes it and insert a predefined decorator on the function definition. The decorated function therefore runs our injected logic whenever it is invoked. Within the decorator, the F2P function is first executed repeatedly, 10 times in a plain loop, as a warm-up, ensuring that the subsequent execution is in a steady state (for example, with lazy imports already absorbed). The decorator then executes the F2P function once more, and, immediately before and after this single invocation, writes a distinctive marker line directly to the standard error stream through a system call. 

Before each execution, we restore the repository to the base commit and apply the ground-truth patch and the test patch, so the F2P test runs in its passing state. We then run the benchmark's single-point command under \texttt{strace}\footnote{\url{https://strace.io}, a tool for tracing system calls.} (in practice, by prepending the strace keyword to the command), and all the system calls produced during the execution are written to a log file. We exclude any instance that cannot sustain in-process repeated execution and therefore fails to collect the required log. For the instances that run through, we analyze the log and, guided by the distinctive marker, precisely locate the window of the last F2P invocation and extract the system calls it produced. Over these calls, we empirically construct a classifier and determine whether the execution involves I/O operations, and exclude any instances that do.

\subsubsection{Filtering Result}
In summary, filtering the raw datasets with the static rule and the dynamic probing yields a corpus of 339 instances on which the rest of the study runs (Table~\ref{tab:funnel}). Among the three filters, the single-test constraint removes the most instances, and SWE-bench-Live loses far more than SWE-bench, consistent with SWE-bench-Live being the harder benchmark indicated by its lower solving rates~\cite{zhang2026swe}. 

\begin{table}[t]
  \centering
  \caption{Corpus construction funnel.}
  \label{tab:funnel}
  \small
  \begin{tabular}{lccc}
    \toprule
 Stage & SWE-bench & SWE-bench-Live & Total \\
    \midrule
 Original dataset             & 500         & 500           & 1{,}000 \\
 w/o a single test            & 155         & 213           & 368 \\
 w/o repeated exec.   & 52          & 35            & 87 \\
 w/o I/O-free exec.      & 63          & 143           & 206 \\
    \midrule
 measurable corpus          & 230 & 109 & 339 \\
    \bottomrule
  \end{tabular}
\end{table}

The repeatability rule removes 87 instances whose failures fall into three classes of increasing strictness. The first class cannot be executed under our reproduction setup. These instances fail to pass their tests stably even under the original suite execution because of unreachable external dependencies or dataset pathologies. The second class, going further, cannot be run as a single-point test. It is worth noting that the F2P test itself is not executed as a single-point test in the original benchmark infrastructure; this native infrastructure runs the complete test files, and a log parser then extracts the test outcome to derive the F2P verdict. Because executing the full test suite is impractical for performance measurement, we combine the native command with the F2P identifier into a single-point execution. It turns out to be feasible in the vast majority of instances, with only 12 failing: some instances' F2P identifier is corrupted, and our reverse derivation still fails to locate them; others fail because they rely on state set by preceding tests in the same file. The third class can be run in single-point but fails under in-process repetition. Their root cause is that each test invocation mutates state that outlives it, and later invocations collide with this residue. The mutated state lives either inside the process, as database rows inserted or caches accumulated, or outside it, as filesystem artifacts and server threads left behind. The collision may surface immediately on the next invocation, or after the accumulated state crosses a threshold. The injected decorator's warm-up loop screens out exactly these instances. 


The I/O-free rule is the second largest source of exclusions, removing 206 instances. We examine the 10 largest contributing repositories and find that they mainly fall into two classes. The first class consists of repositories whose domain is I/O itself, where the tested behavior literally performs file or network operations, such as validating configuration files (\emph{cfn-lint}), rendering PDF documents (\emph{WeasyPrint}), resolving packages (\emph{conan}), extracting media streams (\emph{yt-dlp} and \emph{streamlink}), and scanning infrastructure definitions (\emph{checkov}). The second class consists of general-purpose repositories such as \emph{django}, \emph{sphinx}, \emph{pytest}, and \emph{pylint}; their tested behavior can be pure computation, but the test scaffolding around it usually performs I/O to build up the environment, such as seeding databases, writing build artifacts, or creating temporary files. In these cases, the I/O is a side effect rather than what the test verifies. Because our instrument measures the F2P test as one whole unit, we exclude these instances as well.

Overall, we devise a careful filtering process to ensure the soundness of the analysis that follows. As stringent as the process is, it still leaves a substantial corpus as the data basis for this study.

\subsection{Model Solving and Patch Diversity}
\label{sec:solving}

On the constructed corpus, we employ the widely used \textbf{SWE-agent} framework~\cite{yang2024swe} (the lightweight mini-SWE-agent variant\footnote{\url{https://github.com/SWE-agent/mini-swe-agent}}), together with four state-of-the-art LLMs, \textbf{Claude Opus~5}, \textbf{GPT~5.6}, \textbf{DeepSeek V4 Pro}, and \textbf{Qwen~3.8 Max} (referred to as Claude, GPT, DeepSeek, and Qwen hereafter)\footnote{GPT is the sol version and DeepSeek is the 0813 snapshot. All agent trajectories were produced on August 17--18, 2026.}, to repair each instance under the official execution protocol with a step limit of 250. We select the strongest flagship models each provider currently offers to yield as many correct solution patches from different models as possible, and to keep the subsequent analysis on the current frontier rather than outdated models. Table~\ref{tab:accuracy} reports resolution rates; correctness is graded by the official benchmark evaluation infrastructure. We find that \textbf{Claude} achieves the highest resolution rate on both datasets, while \textbf{DeepSeek} is the lowest. Overall, the correctness ordering is \textbf{Claude $\gg$ Qwen $>$ GPT $\approx$ DeepSeek}. Correctness reflects a model's ability to repair issues, and we examine whether this ability diverges from the runtime performance of the patches the model generates.

\begin{table}[t]
  \centering
  \caption{Resolution rates of the four models on the corpus.}
  \small
  \label{tab:accuracy}
  \begin{tabular}{lccc}
    \toprule
 Model & SWE-bench & SWE-bench-Live & Total \\
    \midrule
 Claude    & \textbf{220 (95.65\%)} & \textbf{76 (69.72\%)} & \textbf{296 (87.32\%)} \\
 Qwen      & 207 (90.00\%) & 75 (68.81\%) & 282 (83.19\%) \\
 GPT       & 208 (90.43\%) & 70 (64.22\%) & 278 (82.01\%) \\
 DeepSeek  & 206 (89.57\%) & 70 (64.22\%) & 276 (81.42\%) \\
    \bottomrule
  \end{tabular}
\end{table}

We then collect all correct patches generated by LLMs and analyze their consistency and diversity. We merge identical patches, since they naturally induce no difference in runtime performance. We define identity by normalized source files rather than raw patches. Specifically, we first apply each patch to the repository at the base commit, discard its modifications to test files following the evaluation protocol~\cite{jimenez2024swe,zhang2026swe}, and sort out the modified files. We parse each modified Python file into an abstract syntax tree, then unparse it back into canonical form, removing comments, docstrings, formatting, and layout differences. We compare non-code files as-is, since no prior evidence suggests they do not affect execution behavior. We treat two patches as the same solution when all their resulting files are equal under this normalization.

\begin{figure}[t]
  \centering
  \includegraphics[width=\linewidth]{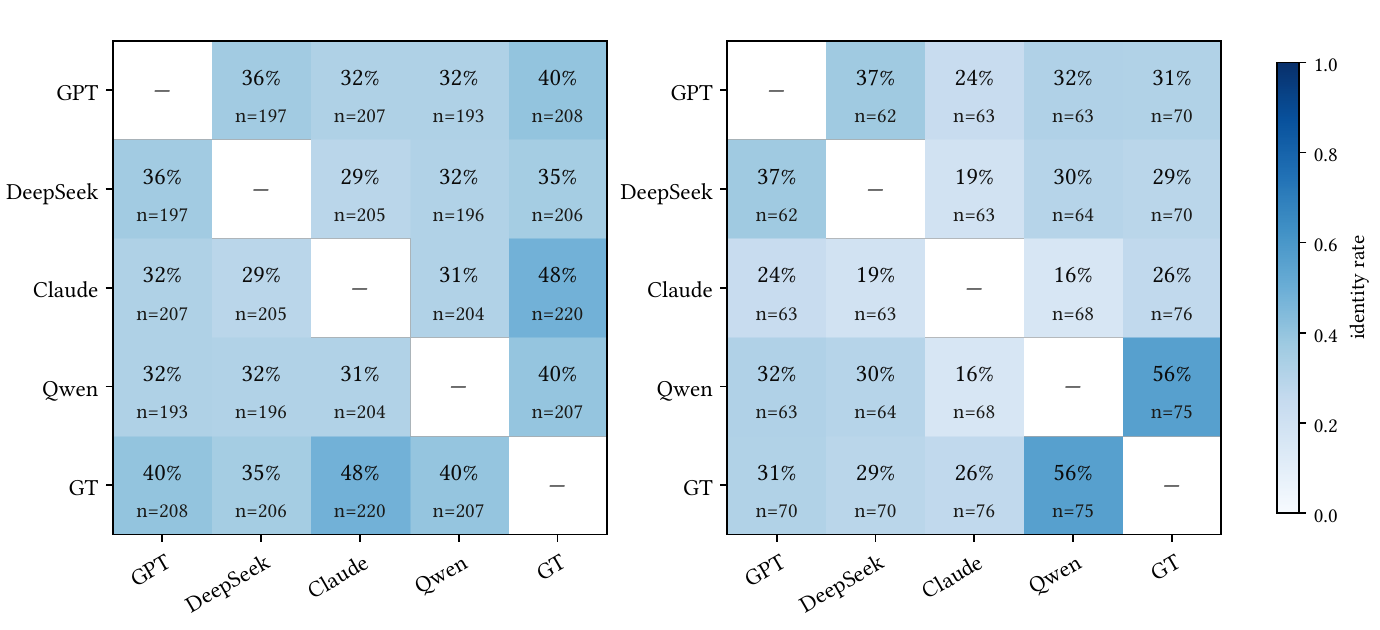}
  \caption{Solution identity between the four models and the ground truth (GT), on the measurable corpus (left: SWE-bench, right: SWE-bench-Live).}
  \label{fig:convergence}
\end{figure}

We include the developer's ground-truth patch as one of the solutions compared for each case. Fig.~\ref{fig:convergence} summarizes the consistency analysis. Each cell corresponds to a pair of sources with the cases that both sources resolved as its base, and reports the fraction of that base on which the two sources submit identical patches. Fig.~\ref{fig:convergence} first shows that patch identity between models is common; every pair of models submits identical patches on a non-negligible fraction of their commonly resolved cases, with GPT and DeepSeek the most consistent pair, at 36\% on SWE-bench and 37\% on SWE-bench-Live. Second, we find that on SWE-bench the model's identity with the ground truth generally exceeds its identity with other models. For example, Claude coincides with the ground truth on 48\% of its resolved cases, well above its identity with any other model; the same holds for GPT and Qwen. In addition, DeepSeek's identity with the ground truth is also closely behind its identity with GPT. Third, on SWE-bench-Live, the models' identity with the ground truth generally remains lower than on SWE-bench. A plausible explanation is that SWE-bench-Live is harder than SWE-bench, as its much lower resolution rates indicate (Table~\ref{tab:accuracy}); its tasks may less often admit the simple, direct solutions like SWE-bench tasks do, so the solution space is possibly larger and the models are less likely to converge on the ground-truth fix. The only exception is Qwen, which matches the ground truth in 56\% of cases, exceeding any source pair.

Table~\ref{tab:diversity} shows the distribution of the merged result. We find that \textbf{262} cases (77.29\%) contain at least two different solutions, and therefore constitute the study corpus, because \textbf{performance comparison is meaningful only when at least two distinct implementations exist}. Table~\ref{tab:diversity} shows that solution diversity is common. In particular, in 11.21\% of cases, all four models produce mutually different patches that also differ from the ground-truth patch, resulting in five distinct solutions in each case. Overall, this diversity provides a solid data basis for the subsequent study of performance differences among different solutions. We also observe that the proportion of cases with at least two different solutions is markedly lower on SWE-bench-Live (67.89\%) than on SWE-bench (81.74\%). We believe that this observation does not contradict the lower identity rates on SWE-bench-Live compared to SWE-bench, as shown in Fig.~\ref{fig:convergence}. In fact, obtaining at least two different solutions requires that at least one model, in addition to the ground truth, produce a correct solution. Therefore, the lower accuracy on SWE-bench-Live (Table~\ref{tab:accuracy}) naturally leads to fewer cases where any model produces a correct solution, resulting in a smaller overall proportion of cases with at least two distinct solutions.

\begin{table}[t]
  \centering
  \caption{The number of distinct solutions after merging identical patches (ground truth included).}
  \label{tab:diversity}
  \small
  \begin{tabular}{lccc}
    \toprule
    \# distinct solutions & SWE-bench & SWE-bench-Live & Total \\
    \midrule
 1 & 42 (18.26\%)  & 35 (32.11\%)  & 77 (22.71\%) \\
 2 & 55 (23.91\%)  & 23 (21.10\%)  & 78 (23.01\%) \\
 3 & 45 (19.57\%)  & 23 (21.10\%)  & 68 (20.06\%) \\
 4 & 60 (26.09\%)  & 18 (16.51\%)  & 78 (23.01\%) \\
 5 & 28 (12.17\%)  & 10 (9.17\%)   & 38 (11.21\%) \\
    \midrule
    $\ge$2 (\textbf{as study corpus}) & \textbf{188 (81.74\%)} & \textbf{74 (67.89\%)} & \textbf{262 (77.29\%)} \\
    \bottomrule
  \end{tabular}
\end{table}
\section{Runtime Performance Evaluation with F2P Tests}
\label{sec:baseline}

We then use the built-in F2P tests to evaluate the runtime performance of solution patches. This section describes the evaluation methodology. We first explain how to instrument each test in \S\ref{sec:baseline-protocol}, then define the metrics and statistical analysis criteria (\S\ref{sec:baseline-analysis}), and finally describe the hardware infrastructure (\S\ref{sec:apparatus}).

\subsection{Execution Protocol}
\label{sec:baseline-protocol}
Each F2P test runs as-is inside its official per-instance Docker environment. The evaluation procedure is organized at two levels~\cite{georges2007statistically}; the inner loop repeats the F2P test function $N$ times within the same process to lift the measured quantity above the noise floor, while the outer loop repeats that process $G$ times as independent replicates to expose run-to-run variation. We evaluate each distinct solution patch of each case using the same procedure. 

\subsubsection{Inner loop.}
We implement the inner loop with the same decoration mechanism as in the filtering phase (\S\ref{sec:filtering}), except that we replace the marked pass with an instrumented loop of $N$ repetitions. A single-point test drives the wrapped test body through a warm-up stage followed by a measurement stage within the same process. During the measurement stage, the F2P function runs $N$ times under instrumentation, and we sum the recorded metrics over the $N$ repetitions to form the measurement for one replicate. To keep total work per measurement roughly uniform across cases for dataset-level aggregation, and to suppress the influence of run-to-run noise, we first anchor $N$ per case during a calibration phase before measurement, letting the measurement of the fastest solution take between 4 and 8 seconds. Technically, we first estimate the time cost of a single execution of the fastest solution with a small number of trial repetitions, derive an initial loop count from that estimate, and then double the count step by step towards the desired $N$ until the test time falls inside the window. The calibrated $N$ is large in practice; it averages $6.8\times10^{5}$ invocations per case over the corpus, with a median of $1.6\times10^{4}$.

\subsubsection{Outer loop.}
The outer execution repeats the whole process $G=5$ times as independent replicates, each with a fresh process. The five per-replicate measurements form the case-level samples for statistical analysis. For each distinct solution, we reset the repository to the base commit, apply the solution patch, run the measurement with both the inner and outer loops, and record the measurement metrics, which we then attribute to the model or the ground truth that produces it. Solutions run under identical execution commands, namely the single-point execution command constructed in \S\ref{sec:probing} but without the \texttt{strace} tool, with the patch as the only difference.

\subsection{Metrics and Statistical Analysis}
\label{sec:baseline-analysis}

\subsubsection{Dual Metrics.}
We record two performance metrics for every measurement. The first is the instruction count, read from hardware performance counters with the \texttt{cirron}\footnote{\url{https://pypi.org/project/cirron/}} profiler. The instruction count is insensitive to core contention and frequency scaling, making it a fine-grained measure of the computation actually performed, adopted in prior work ~\cite{peng2025coffe}. The second is the widely used wall-clock time, which captures end-to-end latency including scheduler, memory, and cache effects that instruction counts do not reflect. It is more prone to visible fluctuations under subtle environmental changes~\cite{peng2025coffe}. The two metrics complement each other, and we report both throughout the paper.

\subsubsection{Discrimination assessment.}
We quantify the magnitude of the performance difference between solutions of each case using RSD following~\cite{peng2025coffe}. Consider a case with $k$ candidate solutions on a specific metric (instruction count or wall-clock time), each solution measured in $G$ replicates, and let $x_{i,g}$ be the measurement of solution $i$ in replicate $g$ ($i=1,\dots,k$, $g=1,\dots,G$). Writing $\bar{x}_i = \frac{1}{G}\sum_g x_{i,g}$ for the mean of solution $i$ and $\bar{x} = \frac{1}{k}\sum_i \bar{x}_i$ for the grand mean over these solution means, RSD is
\begin{equation}
\text{RSD} = \frac{1}{\bar{x}}\sqrt{\frac{1}{k-1}\sum_{i=1}^{k}\big(\bar{x}_i-\bar{x}\big)^2},
\end{equation}
i.e., the between-solution standard deviation (with $k-1$ degrees of freedom) over the between-solution mean.

As this definition shows, RSD captures only the between-solution difference; each solution is represented solely by its replicate mean during the computation, so the within-solution variation never influences the value. We argue that the assessment should exploit every measurement rather than reduce them to a replicate mean. In fact, the RSD-based assessment implicitly assumes that repeated measurements of a solution are fairly stable, an assumption that is fragile in practice, especially when using wall-clock time as the metric. We therefore propose using Welch ANOVA~\cite{welch1951comparison} to assess each case with a statistical significance test. Welch ANOVA is a standard statistical tool, commonly used in software engineering research~\cite{bagherzadeh2021reinforcement,dias2014supporting}. Unlike RSD, Welch ANOVA accounts for within-solution and between-solution variation simultaneously. Another merit is that it generalizes the classical one-way ANOVA to groups with unequal variances, which fits our setting well, since no prior evidence indicates that the case's different solutions share a common variance. 

Specifically, given the measurement $x_{i,g}$ of solution $i$ in replicate $g$, we first apply a logarithmic transformation $y_{i,g}=\ln x_{i,g}$ before statistical testing. This transformation makes differences on the log scale correspond to ratios on the original scale, allowing the analysis to capture relative rather than absolute differences in performance between solutions. The resulting statistic $F_W$ accounts for both between-solution and within-solution variation, with Welch's adjustment for unequal variances across solutions. The $p$-value is
\begin{equation}
p = \Pr\!\left(F_{k-1,\nu} > F_W\right),
\end{equation}
where $F_{k-1,\nu}$ is the $F$ distribution with $k-1$ numerator degrees of freedom and $\nu$ the Satterthwaite approximate denominator degrees of freedom~\cite{welch1951comparison}. The resulting $p$-value tells how likely chance alone would produce a between-solution spread at least as large as the measured one if the case's solutions were in fact equally fast. A small $p$-value therefore means that the measured spread can hardly be attributed to chance, and we take it as evidence of difference. 


To the best of our knowledge, we are the first to apply Welch ANOVA to the performance analysis of different LLM-generated code patches for repository issue resolution. In summary, Welch ANOVA and the $p$-value it produces have four advantages over RSD. First, it weighs the between-solution difference against the within-solution run-to-run variation instead of reporting the former alone. Second, Welch's adjustment allows the within-solution variance to differ across solutions. Third, RSD is vulnerable to over-split solutions. Static comparison shown in \S\ref{sec:solving} may still leave behaviorally identical patches as different candidates, for example, when the same logic is relocated elsewhere (such as moved into a helper function). Such solutions pull the grand mean toward the subgroups they belong to and shrink the between-solution standard deviation, so RSD likely understates the genuine difference. In contrast, Welch ANOVA tests whether all solution means are equal, a hypothesis that does not depend on how many candidates are identical, and its false-positive rate is controlled at the nominal significance level rather than determined by the number of candidates. Lastly, the two statistics differ in how they respond to the number of solutions. The absolute magnitude of RSD is related to the number of solutions compared in a given case, making it difficult to define a universal RSD threshold across cases. In contrast, Welch ANOVA accounts for the number of groups through the test statistic $F_W$ and the degrees of freedom in $F_{k-1,\nu}$, so each case statistic is compared against its matching reference distribution. As a result, it yields a $p$-value on a common statistical scale, allowing the same significance level to be applied across cases. Following conventional practice~\cite{fisher1934statistical,wasserstein2016asa}, we regard a case as exhibiting statistically distinguishable performance on a metric when $p < 0.05$. In the following experiments, we compute RSD and $p$-value separately for instruction count and wall-clock time.

\subsection{Hardware Infrastructure}
\label{sec:apparatus}

All measurements share the same hardware infrastructure. All run on a server with an Intel Xeon 6982P-C processor (32 physical cores, each with 2 logical cores, 128\,GB memory), and each execution runs inside a Docker container on this server. To measure the execution performance precisely, we devise a careful hardware scheduling policy, most of which follows~\cite{ma2026swefficiency}, while ours is stricter.

\subsubsection{Core reservation.}
We first identify the server's CPU topology, including the physical cores, their hyperthread siblings, and the NUMA (Non-Uniform Memory Access) nodes. We reserve 2 physical cores for the Docker daemon and 1 core for the operating system, so background activity such as container and image management (fetching images, launching a container, or cleaning it up afterwards) is confined to these reserved CPUs and cannot steal cycles from the measurement cores. As a result, we can parallelize work across multiple groups without coupling their performance to background activity.

\subsubsection{Kernel-level CPU isolation.}
We then partition the remaining physical cores into disjoint groups of 4; each group serves 1 measured case. For each group, we create a cgroup v2\footnote{\url{https://docs.kernel.org/admin-guide/cgroup-v2.html}} isolated cpuset partition whose cpuset covers all logical cores of the 4 physical cores, including their hyperthread siblings (8 logical cores total). The partition reclaims these cores from the host scheduler, and processes outside the partition subtree, including system services and other containers, cannot be scheduled on them. This isolation is stronger than in~\cite {ma2026swefficiency}, where containers are merely pinned to specific cores by affinity. However, this strategy constrains only the pinned processes. It cannot ensure that the container exclusively occupies the chosen cores; other processes may still land on them, especially on a multi-user server, making the measurement unstable and unreliable.

Each case's container is then started and migrated into a leaf cgroup under this partition, whose cpuset holds only 1 logical core per physical core. The siblings stay idle inside the partition and are unusable by the host or the container, so no logical cores in the same group share a physical core. This grouping strategy improves repeatability, since each physical core's private caches are used only by the single logical core allocated to it, and other cores never contend for the same cache, whether within a group or across groups. After the reservations, $(32-2-1)/4 = 7$ disjoint groups fit on the server, so at most 7 cases are measured in parallel.

\subsubsection{Memory limits and NUMA binding.}
Each container is bound to the NUMA memory node of its allocated cores. The server exposes a single NUMA node, so this binding keeps all memory local and prevents cross-node access. Each container's memory is capped at 4\,GB, and swap is disabled, which bounds the per-case footprint, keeps memory allocation predictable, and removes the disk traffic that paging would introduce. These constraints complement the CPU isolation described above and further enhance the stability and reproducibility of the measurement.

\section{Results and Analysis}
\label{sec:baseline-results}
With the corpus and the protocol in place, we report what the F2P tests deliver. We assess whether the F2P tests distinguish the cases' solutions (\S\ref{sec:disc_prop}) and compare the models' performance (\S\ref{sec:perf_analy}).

\begin{table}[t]
  \centering
  \caption{Number and share of distinguishable cases ($p<0.05$) with F2P tests.}
  \small
  \label{tab:baseline-dist}
  \begin{tabular}{lcc}
    \toprule
 Dataset & Instructions & Wall time \\
    \midrule
 All (n=262)   & 110 (41.98\%) & 91 (34.73\%) \\
 SWE-bench  (n=188)   & 80 (42.55\%)  & 65 (34.57\%) \\
 SWE-bench-Live  (n=74)   & 30 (40.54\%)   & 26 (35.14\%) \\
    \bottomrule
  \end{tabular}
\end{table}

\begin{table}[t]
  \centering
  \caption{Per-case between-solution RSD with F2P tests.}
  \label{tab:baseline-rsd}
  \small
  \begin{tabular}{l ccc ccc}
    \toprule
    & \multicolumn{3}{c}{Instructions} & \multicolumn{3}{c}{Wall time} \\
    \cmidrule(lr){2-4} \cmidrule(lr){5-7}
 Dataset & median & max & mean & median & max & mean \\
    \midrule
 All (n=262)  & 0.21\% & 132.68\% & 3.13\% & 0.59\% & 131.87\% & 3.49\% \\
 SWE-bench  (n=188)  & 0.27\% & 132.68\% & 3.71\% & 0.58\% & 131.87\% & 4.10\% \\
 SWE-bench-Live  (n=74)  & 0.15\% & 45.22\%  & 1.68\% & 0.62\% & 42.55\%  & 1.97\% \\
    \bottomrule
  \end{tabular}
\end{table}

\subsection{Test Discriminative Property}
\label{sec:disc_prop}
We first examine the statistical distinguishability between solutions with the F2P tests. Table~\ref{tab:baseline-dist} shows that, under the F2P tests, nearly half of the cases exhibit a statistically distinguishable difference between various fixes. This indicates that performance differences between model patches are fairly common. We also find that instruction count is more discriminative than wall-clock time, which is reasonable since wall-clock time is noisier than instruction count~\cite{peng2025coffe}; its within-solution variation is larger, and the resulting $p$-values are correspondingly inflated. In addition, we find that SWE-bench and SWE-bench-Live have similar shares of distinguishable cases.

We then examine solution performance spread by RSD. Table~\ref{tab:baseline-rsd} shows that the median and mean for both metrics remain low, and the overall distribution exhibits a clear heavy tail. In particular, on SWE-bench the between-solution difference of some cases reaches as high as 132.68\%, indicating an even more pronounced heavy tail in this subset. We further examine how RSD is distributed across cases with different numbers of distinct solutions and find the distributions differ markedly, so no unified RSD threshold applies across them.


\begin{figure}[t]
  \centering
  \includegraphics[width=\linewidth]{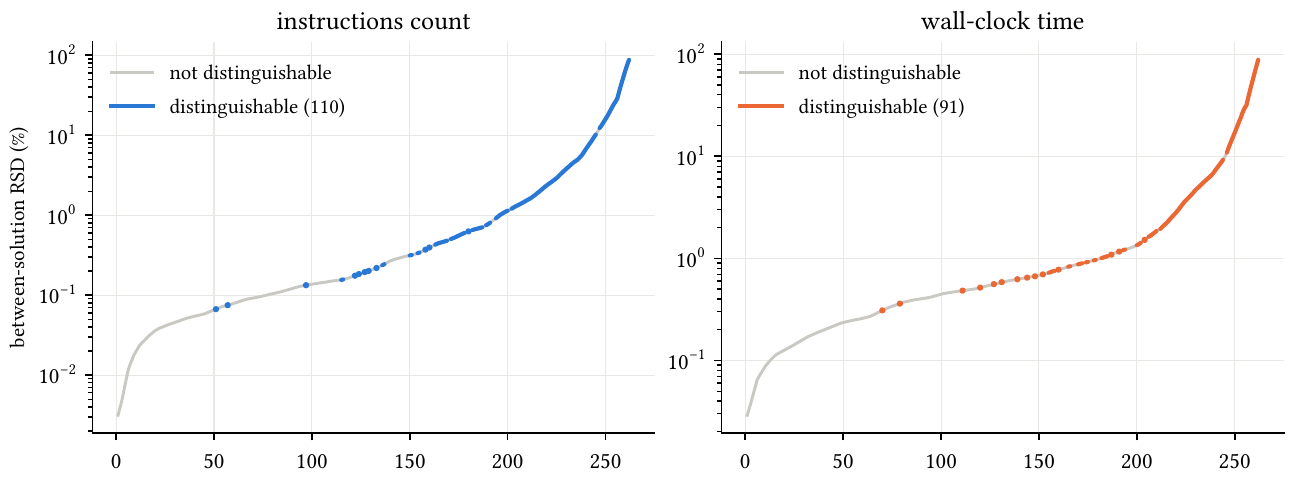}
  \caption{Per-case between-solution RSD with F2P tests, sorted ascending. The gray curve shows RSD for all cases; colored segments and dots mark cases with $p<0.05$.}
  \label{fig:rsd-p-sorted}
\end{figure}

To analyze the difference between the $p$-value and RSD, we further compare them case by case, as shown in Fig.~\ref{fig:rsd-p-sorted}. Specifically, we first take the between-solution RSD for each case, sort the values in ascending order, and use the sorted curve as the figure base. We then overlay, with colored points, the cases whose $p$-values fall below the threshold and are therefore classified as distinguishable. Clearly, when RSD stays high, toward the right end of the curve, the curve is almost continuously covered by colored points, showing that the two statistics render largely consistent verdicts that the solutions of these cases differ markedly. However, as RSD slides from high to low toward the left end, isolated colored points begin to appear on the left part of the curve, showing that Welch ANOVA still classifies these cases as distinguishable even when their RSD is relatively low. Overall, the disagreement between the two statistics gradually emerges as RSD decreases.

To examine this phenomenon further, we pick an isolated point from the left end of the curve. One case is \emph{pydata\_\_xarray-9682} from SWE-bench-Live, whose RSD on instruction count is only 0.075\% while its $p$-value is 0.002, so Welch ANOVA still regards it as distinguishable despite the low RSD. In this case, the four models and the ground truth produced five solutions that differed in form to varying degrees. Our inspection reveals that, in terms of the core resolution idea, these five solutions fall into two groups: group A, consisting of the patches from the ground truth, GPT, and Claude, and group B, consisting of those from DeepSeek and Qwen. These groups mainly adopt two different strategies to solve the same issue. We then pair every source in group A with every source in group B and compute the $p$-value for each pair. All six cross-group pairs obtain $p < 0.05$, while all four within-group pairs obtain $p$ far above 0.05. The $p$-value captures the genuine performance difference between two groups, while RSD fails. In addition, at the upper right of the curve, a single point with a high RSD is uncolored. We confirm that its spread stems from run-to-run noise within the solutions due to the test's randomness, rather than any genuine difference between them.


Note that we have carefully merged different solutions. Existing work, by contrast, does not merge solutions before comparing runtime performance using RSD alone, so we argue that prior work may overlook differences that actually exist between solutions~\cite{peng2025coffe}. Considering that prior work judges the quality of test scripts by RSD~\cite{peng2025coffe}, namely that, on a given solution set, a higher RSD indicates a more effective and discriminative test instrument, we call on future work also to include Welch ANOVA, since it captures the genuine differences that RSD fails to reveal. In conclusion, we summarize our findings as follows:

\smallskip
\noindent\textbf{Finding 1.1:} The built-in F2P tests have meaningful discriminating power; in nearly half of the cases, different patches show statistically distinguishable differences, and instruction count is more discriminating than wall-clock time.

\smallskip
\noindent\textbf{Finding 1.2:} Welch ANOVA and RSD do not always render the same verdict. The former is more robust, while the latter is likely to overlook genuine runtime differences and retain cases without meaningful distinctions.

\subsection{Model Performance Analysis}
\label{sec:perf_analy}
We now turn to what the measurements reveal about the models. We first study the speedup through three comparisons: each model on the cases it solved, the four models together on their common solved subset, and every model pair head-to-head. Beyond speedup, we also examine which model produces the largest number of fastest solutions across all cases.

\subsubsection{Speedup Comparisons}
\label{sec:accel}
The three speedup comparisons rest on a single way of pooling per-case performance observations into one number per model. For the two sources under comparison, $i$ and $j$ (either two models, or a model and the ground truth), we compute the per-case speedup ratio of $i$ relative to $j$ as $r_{i,j}=\bar{x}_i/\bar{x}_j$, where $\bar{x}_i$ denotes the mean measurement on the solution produced by source $i$ and $r<1$ means source $i$ is faster. The speedup is independent of each case's inner loop: the different solutions run the same number of invocations, so the workload scale cancels out.


After that, we collect the ratios of the $n$ covered cases into $\mathbf{r}=(r_1,\dots,r_n)$ and aggregate them as a \textbf{geometric mean}, which is commonly used for performance evaluation~\cite{sehgal2026formulacode,fleming1986not,citron2006harmonic}:
\begin{equation}
G(\mathbf{r})=\Bigl(\prod_{c=1}^{n} r_c\Bigr)^{1/n}.
\end{equation}
A characteristic of the geometric mean is its symmetry between speedups and slowdowns. For example, given two patch sources $A$ and $B$, $A$ runs twice as fast as $B$ in one case and twice as slow as $B$ in another. The geometric mean neutralizes them to exactly $1$. This characteristic further gives rise to an advantage: the geometric mean keeps the conclusion stable when the numerator and denominator are swapped. For ratios $\mathbf{x}=(x_1,\dots,x_n)$ and $\mathbf{y}=(y_1,\dots,y_n)$ with $x_i y_i=1$ for every $i$,
\begin{equation}
G(\mathbf{y})=\Bigl(\prod_i y_i\Bigr)^{1/n}=\Bigl(\prod_i x_i^{-1}\Bigr)^{1/n}=\Bigl(\prod_i x_i\Bigr)^{-1/n}=G(\mathbf{x})^{-1}.
\end{equation}
Reconsider the two sources $A$ and $B$ mentioned above: the geometric mean gives an even score between $A$ and $B$, regardless of which source sits in the denominator. The arithmetic mean, for example, reports $1.25$ on $\mathbf{x}=(1/2,\,2)$ with $A$ against $B$, and again $1.25$ on $\mathbf{y}=(2,\,1/2)$ with $B$ against $A$, so the ordering of $A$ and $B$ comes out contradictory and can never be settled.
Note that several related works~\cite{ma2026swefficiency,shetty2026gso,smith1988characterizing} rely on the harmonic mean rather than the geometric one, since they usually aggregate speedup against a fixed expert reference. However, in our setting, the ground truth cannot be assumed to be the most efficient; it is one solution of equal standing among others, so pairwise orientations are interchangeable, making the geometric mean more suitable.

All speedup comparisons below are reported in two settings: over the full set of cases, and over the distinguishable subset of each metric (110 cases for instruction count and 91 for wall-clock time, as shown in Table~\ref{tab:baseline-dist}). The former is an unconditional, corpus-level view: no cases are pre-selected, and the aggregate reflects overall speedup. The latter deliberately amplifies differences by dropping cases where the tests detect no difference between solutions, so the separation between sources emerges. Reporting both allows a direct comparison between the two settings.

\begin{table}[t]
  \centering
  \caption{Per-model comparison with the ground-truth. A win (loss) means the model patch is faster
 (slower) than the ground-truth patch, and a tie means the
 two patches are identical. Speedup is the geometric mean of
 per-case ratios (lower is faster). Scored gives the number of cases the
 model resolved, on all 262 cases in (a), and on that metric's
 distinguishable cases (110 for instructions, 91 for wall time) in (b).}
  \label{tab:baseline-scorecard}
  \begin{subtable}[t]{\textwidth}
    \centering\small
    \caption{All cases}
    \label{tab:baseline-scorecard-all}
    \begin{tabular}{l c c ccc ccc}
      \toprule
      & & & \multicolumn{3}{c}{Instructions} & \multicolumn{3}{c}{Wall time} \\
      \cmidrule(lr){4-6} \cmidrule(lr){7-9}
 Model & Scored & Tie & Win & Loss & Accel. & Win & Loss & Accel. \\
      \midrule
 GPT       & 237/262 & 64 & 95  & 78 & \textbf{99.00\%}  & 89  & 84 & 99.05\% \\
 DeepSeek  & 240/262 & 57 & 100 & 83 & 99.10\% & 108 & 75 & \textbf{98.99\%} \\
 Claude    & \textbf{252/262} & 82 & 84  & 86 & 99.36\% & 81  & 89 & 99.39\% \\
 Qwen      & 245/262 & 87 & 78  & 80 & 100.51\% & 80  & 78 & 100.46\% \\
      \bottomrule
    \end{tabular}
  \end{subtable}

  \vspace{4pt}
  \begin{subtable}[t]{\textwidth}
    \centering\small
    \caption{Distinguishable cases}
    \label{tab:baseline-scorecard-disc}
    \begin{tabular}{l cc ccc cc ccc}
      \toprule
      & \multicolumn{5}{c}{Instructions} & \multicolumn{5}{c}{Wall time} \\
      \cmidrule(lr){2-6} \cmidrule(lr){7-11}
 Model & Scored & Tie & Win & Loss & Accel. & Scored & Tie & Win & Loss & Accel. \\
      \midrule
 GPT       & 100/110 & 20 & 43 & 37 & \textbf{97.67\%}  & 86/91 & 15 & 40 & 31 & 97.42\% \\
 DeepSeek  & 101/110 & 20 & 44 & 37 & 97.94\% & 84/91 & 16 & 40 & 28 & \textbf{97.27\%} \\
 Claude    & \textbf{109/110} & 29 & 36 & 44 & 98.67\% & \textbf{89/91} & 18 & 30 & 41 & 98.54\% \\
 Qwen      & 101/110 & 32 & 31 & 38 & 101.25\% & 83/91 & 27 & 27 & 29 & 101.40\% \\
      \bottomrule
    \end{tabular}
  \end{subtable}
\end{table}

\paragraph{Comparison on Each Model's Resolved Cases.}
We first examine each model's runtime performance on the cases it resolved, comparing each model against the ground truth via the speedup (Table~\ref{tab:baseline-scorecard}). Note that the models differ in the cases they resolved, so the comparison bases are not identical. Across the Scored columns, Claude attains the highest resolution rate throughout, on the full set and on the distinguishable subsets of both metrics. In terms of runtime performance, GPT, DeepSeek, and Claude all run faster than the ground truth on their own resolved cases, on the full set and the distinguishable subsets alike, while Qwen is consistently slower. Besides, for all models, the gaps to the ground truth become more pronounced on the distinguishable subsets (Table~\ref{tab:baseline-scorecard-disc}). On the win and loss counts, GPT and DeepSeek always win more cases than they lose against the ground truth in every data category; Claude always loses more cases than it wins even though its aggregate stays faster, which indicates that its winning cases speed up by a larger margin than its losing cases slow down; Qwen keeps its wins and losses roughly balanced, suggesting that its losing cases slow down by a larger margin than its winning cases speed up. Moreover, for GPT and DeepSeek, the win and loss counts far exceed the tie counts, with Claude in between, while for Qwen the three counts stay roughly equal, suggesting that Qwen and Claude tend to produce patches identical to the ground truth, consistent with the identity analysis in Fig.~\ref{fig:convergence}.

\begin{table}[t]
  \centering
  \caption{Model speedup against the ground truth on the commonly solved cases, aggregated as the geometric mean. All means the full common subset (n=211), while Distinguishable intersects it with each metric's distinguishable cases (n=90 for instructions, n=75 for wall time).}
  \label{tab:baseline-models}
  \small
  \begin{tabular}{l cc cc}
    \toprule
    & \multicolumn{2}{c}{Instructions} & \multicolumn{2}{c}{Wall time} \\
    \cmidrule(lr){2-3} \cmidrule(lr){4-5}
 Model & All & Distinguishable & All & Distinguishable \\
    \midrule
 GPT         & \textbf{98.93\%} & \textbf{97.53\%} & 99.01\% & 97.23\% \\
 DeepSeek    & 99.07\% & 97.89\% & \textbf{98.97\%} & \textbf{97.21\%} \\
 Claude      & 99.33\% & 98.42\% & 99.33\% & 98.18\% \\
 Qwen        & 100.71\% & 101.68\% & 100.71\% & 102.00\% \\
    \bottomrule
  \end{tabular}
\end{table}

\paragraph{Comparison on the Commonly Resolved Cases.}
In Table~\ref{tab:baseline-scorecard}, the models solved different cases, so their performances are not directly comparable. We therefore select the subset of cases all four models solved and compute each model's speedup against the ground truth on this subset, making the models' and the ground truth's performance all comparable (Table~\ref{tab:baseline-models}). The analysis covers both the full set and the distinguishable subsets for both metrics. GPT and DeepSeek clearly lead the field, with the former ahead on instruction count and the latter on wall-clock time; Claude trails the two yet remains faster than the ground truth, while Qwen performs worst and is even slower than the ground truth overall. Taken together, the runtime performance of the models exhibits the ordering \textbf{GPT $\approx$ DeepSeek $>$ Claude $\gg$ Qwen}, which stands far from their resolution rate ordering \textbf{Claude $\gg$ Qwen $>$ GPT $\approx$ DeepSeek} reported in Table~\ref{tab:accuracy}. The runtime performance of the produced patches and the models' resolution rates diverge markedly rather than coincide, so we call for future work to account for patch performance in addition to resolution rate. Another clear finding is that the model ordering stays strictly consistent between the full set and the distinguishable subsets for both metrics, and the gaps between models widen further on the distinguishable subsets. These phenomena indicate that the distinguishable subsets produced by Welch ANOVA stand for the full set well, removing non-discriminating cases that flatten differences between models.

\begin{table}[t]
  \centering
  \caption{Pairwise head-to-head speedup with F2P tests. In each pair, model A is faster on instruction count. Speedup aggregates the per-case A/B ratios. Win (loss) means A is faster (slower), and Tie means identical. Total counts the commonly solved cases in (a) and the distinguishable ones among them in (b).}
  \label{tab:baseline-pairwise}
  \begin{subtable}[t]{\textwidth}
    \centering\small
    \caption{All cases}
    \label{tab:baseline-pairwise-all}
    \begin{tabular}{l c c ccc ccc}
      \toprule
      & & & \multicolumn{3}{c}{Instructions} & \multicolumn{3}{c}{Wall time} \\
      \cmidrule(lr){4-6} \cmidrule(lr){7-9}
 A: B & Total & Tie & Win & Loss & Accel. & Win & Loss & Accel. \\
      \midrule
 GPT: Qwen          & 223 & 49 & 92  & 82  & \textbf{98.35\%} & 91  & 83 & 98.42\% \\
 DeepSeek: Qwen     & 229 & 50 & 94  & 85  & 98.49\% & 94  & 85 & \textbf{98.40\%} \\
 Claude: Qwen       & 236 & 39 & 95  & 102 & 98.74\% & 99  & 98 & 98.77\% \\
 GPT: Claude        & 231 & 43 & 108 & 80  & 99.61\% & 107 & 81 & 99.65\% \\
 DeepSeek: Claude   & 233 & 36 & 111 & 86  & 99.72\% & 110 & 87 & 99.59\% \\
 GPT: DeepSeek      & 225 & 60 & 82  & 83  & 99.87\% & 76  & 89 & 100.05\% \\
      \bottomrule
    \end{tabular}
  \end{subtable}

  \vspace{4pt}
  \begin{subtable}[t]{\textwidth}
    \centering\small
    \caption{Distinguishable cases}
    \label{tab:baseline-pairwise-disc}
    \begin{tabular}{l cc ccc cc ccc}
      \toprule
      & \multicolumn{5}{c}{Instructions} & \multicolumn{5}{c}{Wall time} \\
      \cmidrule(lr){2-6} \cmidrule(lr){7-11}
 A: B & Total & Tie & Win & Loss & Accel. & Total & Tie & Win & Loss & Accel. \\
      \midrule
 GPT: Qwen          & 92  & 17 & 45 & 30 & \textbf{96.07\%} & 78 & 14 & 36 & 28 & 95.56\% \\
 DeepSeek: Qwen     & 95  & 18 & 43 & 34 & 96.46\% & 78 & 16 & 30 & 32 & \textbf{95.48\%} \\
 Claude: Qwen       & 100 & 11 & 45 & 44 & 97.05\% & 81 & 8  & 34 & 39 & 96.51\% \\
 GPT: Claude        & 100 & 10 & 55 & 35 & 99.14\% & 85 & 9  & 48 & 28 & 99.02\% \\
 DeepSeek: Claude   & 101 & 10 & 56 & 35 & 99.41\% & 83 & 9  & 45 & 29 & 98.91\% \\
 GPT: DeepSeek      & 95  & 18 & 38 & 39 & 99.67\% & 82 & 15 & 33 & 34 & 100.04\% \\
      \bottomrule
    \end{tabular}
  \end{subtable}
\end{table}

\paragraph{Pairwise Comparison across Models.}
We further compare the models pairwise through the speedup on the cases both models solved. The geometric mean aggregates the speedup, so A: B equals B: A with the inverse value in Table~\ref{tab:baseline-pairwise}. This speedup computation is equivalent to dropping the ground truth as the link, which relies on the transitivity of the speedup preserved by the geometric mean. Also, as noted earlier, the geometric mean allows the numerator and denominator to swap, so we orient each pair to keep the instruction-count speedup below 1, meaning the chosen A is always faster than B to present the comparison more clearly. We find that GPT and DeepSeek both lead the other two models; the largest instruction-count gap is GPT over Qwen, while the largest wall-clock gap is DeepSeek over Qwen, consistent with their leading positions in Table~\ref{tab:baseline-models}. GPT and DeepSeek are comparable, with the speedup hovering around parity on both metrics. Overall, the distinguishable subsets agree with the full set on the pairwise speedup. On the case counts, wins outnumber losses in almost every model pair, consistent with the speedup. The exception is the Claude-Qwen pair, where Claude wins fewer cases than it loses yet still runs faster than Qwen overall. In addition, tie counts between models are far smaller than win and loss counts, in contrast to the frequent identity with the ground truth in Table~\ref{tab:baseline-scorecard}. 


\subsubsection{Champion Distribution}
\label{sec:champion}
The speedup analyses above show that the overall speedup does not fully agree with the case counts, so we further examine the champion distribution over all cases (Fig.~\ref{fig:champion-shipped}). Each case casts one vote for the source that provides the fastest solution, and the vote is split evenly among the models and the ground truth that produced the same fastest one. The comparison covers the distinguishable subset of each metric. On the full set, DeepSeek produces the fastest patches in most cases on both instruction count and wall-clock time, whereas Claude produces the fewest on both. In the number of fastest solutions, both instruction count and wall-clock time basically keep the ordering \textbf{DeepSeek $>$ GPT $\gg$ Qwen $>$ Claude}. On the distinguishable subset of each metric, the gaps widen further: GPT and DeepSeek still lead, while Qwen and Claude remain at a relatively low level, with Qwen slightly above Claude. Combining the preceding analyses of the model speedup in \S\ref{sec:accel} and the champion distribution, we reach the following findings:

\begin{figure*}[t]
  \centering
  \includegraphics[width=\textwidth]{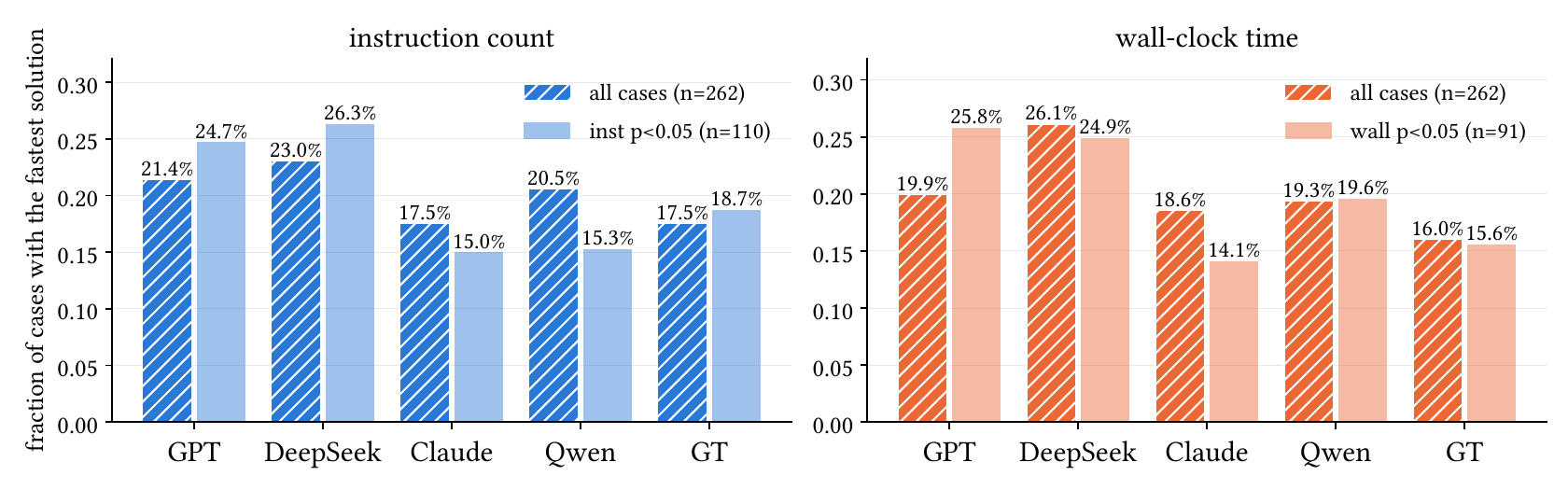}
  \caption{Champion distribution with F2P tests. Each bar shows the share of cases whose fastest solution comes from each source, with the score split evenly among the sources that produced it.}
    \label{fig:champion-shipped}
\end{figure*}

\smallskip
\noindent\textbf{Finding 2.1:} Resolution rates and runtime performance appear as two independent dimensions of model capability, so the former alone cannot substitute for evaluating the patches.

\smallskip

\noindent\textbf{Finding 2.2:} The distinguishable subsets are highly consistent with the full set in model comparisons, and the gaps between models widen on the distinguishable ones.

\smallskip

\noindent\textbf{Finding 2.3:} The models' aggregate speedup generally agrees with how often they produce the fastest patch solution, though the agreement is not complete.


\subsection{Case Study}
\label{sec:case_study}
\begin{figure}[t]
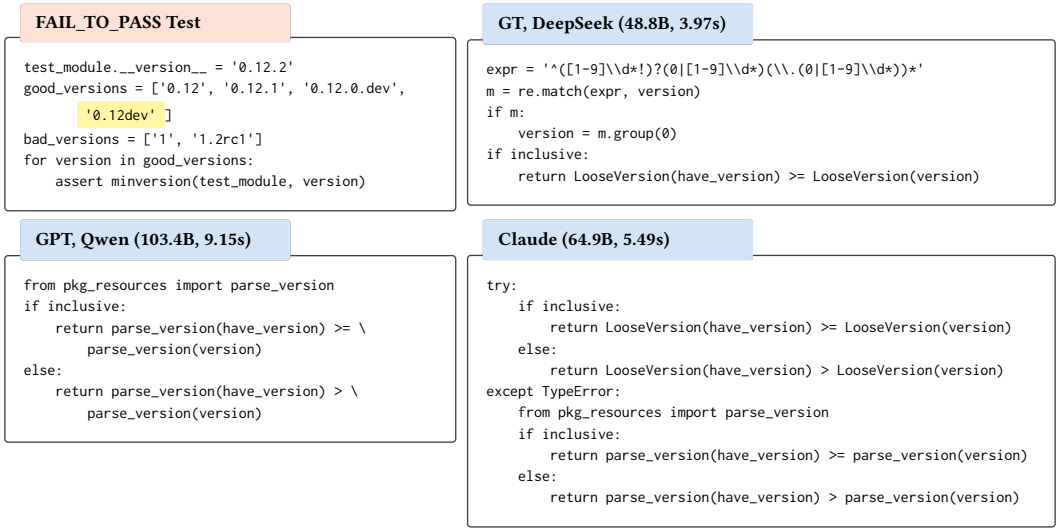

  \centering\tiny
  \begin{minipage}[t]{0.43\textwidth}
    \vspace{0pt}
    \begin{codecard}{FAIL\_TO\_PASS Test}
    \begin{lstlisting}[basicstyle=\tiny\ttfamily, columns=fullflexible,
 keepspaces=true, breaklines=true, aboveskip=0pt, belowskip=0pt,
 escapeinside={(*}{*)}]
test_module.__version__ = '0.12.2'
good_versions = ['0.12', '0.12.1', '0.12.0.dev', (*\colorbox{yellow!45}{\tiny\ttfamily\textquotesingle 0.12dev\textquotesingle}*)]
bad_versions = ['1', '1.2rc1']
for version in good_versions:
    assert minversion(test_module, version)
\end{lstlisting}
    \end{codecard}
  \end{minipage}\hfill
  \begin{minipage}[t]{0.56\textwidth}
    \vspace{0pt}
    \begin{codecard}[champblue!20]{GT, DeepSeek (48.8B, 3.97s)}
    \begin{lstlisting}[basicstyle=\tiny\ttfamily, columns=fullflexible, keepspaces=true, breaklines=true, aboveskip=0pt, belowskip=0pt]
expr = '^([1-9]\\d*!)?(0|[1-9]\\d*)(\\.(0|[1-9]\\d*))*'
m = re.match(expr, version)
if m:
    version = m.group(0)
if inclusive:
    return LooseVersion(have_version) >= LooseVersion(version)
\end{lstlisting}
    \end{codecard}
  \end{minipage}

  \smallskip
  \begin{minipage}[t]{0.43\textwidth}
    \vspace{0pt}
    \begin{codecard}[champblue!20]{GPT, Qwen (103.4B, 9.15s)}
    \begin{lstlisting}[basicstyle=\tiny\ttfamily, columns=fullflexible,
 keepspaces=true, breaklines=true, aboveskip=0pt, belowskip=0pt]
from pkg_resources import parse_version
if inclusive:
    return parse_version(have_version) >= \
        parse_version(version)
else:
    return parse_version(have_version) > \
        parse_version(version)
\end{lstlisting}
    \end{codecard}
  \end{minipage}\hfill
  \begin{minipage}[t]{0.56\textwidth}
    \vspace{0pt}
    \begin{codecard}[champblue!20]{Claude (64.9B, 5.49s)}
    \begin{lstlisting}[basicstyle=\tiny\ttfamily, columns=fullflexible,
 keepspaces=true, breaklines=true, aboveskip=0pt, belowskip=0pt]
try:
    if inclusive:
        return LooseVersion(have_version) >= LooseVersion(version)
    else:
        return LooseVersion(have_version) > LooseVersion(version)
except TypeError:
    from pkg_resources import parse_version
    if inclusive:
        return parse_version(have_version) >= parse_version(version)
    else:
        return parse_version(have_version) > parse_version(version)
\end{lstlisting}
    \end{codecard}
  \end{minipage}
  \caption{The F2P test and the three repair strategies for \emph{astropy\_\_astropy-7671}, with their instruction counts and wall-clock times. The test patch introduces the highlighted code.}
  \label{fig:case-7671}
\end{figure}

Aggregate statistics show the frequency and magnitude of performance differences among solutions, but they do not explain what those differences are. We therefore dissect one case in depth, namely \emph{astropy\_\_astropy-7671} from SWE-bench. The case shown in Fig.~\ref{fig:case-7671} presents the last step of a story spanning two PRs\footnote{\url{https://github.com/astropy/astropy/pull/7647}; \url{https://github.com/astropy/astropy/pull/7671}} related to the \emph{minversion} function in the \emph{astropy} repository. The original implementation of \emph{minversion} mainly parsed versions with the \emph{pkg\_resources} parser. It fell back on \emph{LooseVersion} when \emph{pkg\_resources} was unavailable, and carried a \textbf{regular-expression sanitization} that truncated a version string to its numeric prefix before comparing it with \emph{LooseVersion}. 

Initially, PR 7647 dropped \textbf{heavy} \emph{pkg\_resources} in favor of the \textbf{light} \emph{LooseVersion} alone, halving the package import time from 0.28 to 0.16 seconds. During its review, a sanity check showed that \emph{LooseVersion} successfully compares \emph{3.0dev0} with \emph{3.0dev1}, so \textbf{the sanitizer was carelessly judged obsolete and removed} in the same PR. The sanity check, however, did not cover the scenario where one version's numeric components are compared with another's string components, which the sanitizer had previously guarded against. The repository's test suite also did not cover this situation. The careless removal therefore led to an error, so PR 7671 restored the sanitizer verbatim. The test in PR 7671 appends \emph{0.12dev}, shown in the first card of Fig.~\ref{fig:case-7671}, to compare against an installed version of \emph{0.12.2} to expose this scenario. Semantically, \emph{0.12dev} is a development pre-release version of \emph{0.12} and therefore precedes \emph{0.12.2}; the sanitizer realizes the comparison by truncating \emph{0.12dev} to \emph{0.12}. Without it, \emph{LooseVersion} raises a \emph{TypeError} when comparing the last components of \emph{0.12dev} and \emph{0.12.2}, namely \emph{dev} and \emph{2}. We refer readers to the two PRs for further details.

We now examine the three repair strategies the five sources produced, shown in the remaining cards of Fig.~\ref{fig:case-7671}. The ground truth and DeepSeek restore the sanitizer and stay on the lightweight \emph{LooseVersion} parser. GPT and Qwen switch to the heavy \emph{pkg\_resources} parser unconditionally, while Claude keeps \emph{LooseVersion} and falls back on \emph{pkg\_resources} when the former raises an error. Every patch passes the F2P test, but performance measurements separate them: the three strategies cost 48.8/103.4/64.9 billion instructions and 3.97/9.15/5.49 seconds, respectively. Overall, DeepSeek preserves the developer's initial intent to remove the heavy dependency, while the others silently reintroduce it. The correctness tests cannot detect this architectural regression, which our performance analysis successfully reveals.

\section{Threats to Validity}
\label{sec:threats}
\textbf{Internal validity.}
First, we use the F2P test as the unit of execution and assume the data-preparation and assertion phases cost the same across all correct solutions for a case. However, some patches may occasionally alter the cost of these two phases, contaminating the performance signal. Our method cannot detect this contamination. Second, necessary repeated executions of F2P tests introduce another possibility: only a fraction of the invocations truly execute the repair-relevant code, while later ones merely read the state left behind, so measurements degenerate to indistinguishable tiny values, reducing the discrimination among solutions. We acknowledge these limitations.

\noindent\textbf{External validity.}
Our findings depend on the corpus scope. First, we filter the benchmarks down to a measurable corpus. The filtering is restrictive but entirely deterministic, built on reproducible static rules and dynamic analysis, and should be generalizable. Second, both benchmarks draw from Python repositories. In future work, we will extend the measurement to languages more sensitive to performance, such as C and C++. Third, all solutions come from a single SWE-agent framework, so patch diversity stems mainly from the models rather than the agent framework. Nevertheless, our analysis covers human-written ground-truth patches, which broadens the comparison. We plan to include more frameworks, such as OpenHands.
\section{Related Work}
\label{sec:related}

\paragraph{Repository-level issue resolution.}
SWE-bench packages real issues from real-world repositories, and resolving them demands long-horizon reasoning~\cite{jimenez2024swe}; SWE-bench-Live refreshes the pool with post-cutoff pull requests to reduce contamination~\cite{zhang2026swe}. Agentic systems supply the machinery, from SWE-agent's agent-computer interface~\cite{yang2024swe} to the OpenHands platform~\cite{wang2025openhands}. The stack closes with a boolean verdict on whether the patch passes. We build on them and measure the runtime behavior of patches the verdict discards, using the benchmarks' built-in tests.

\paragraph{Code performance optimization.}
One line of work optimizes the performance of already correct code. They learn optimization diffs for C/C++~\cite{chen2024supersonic}, refine generated code with runtime feedback~\cite{peng2025perfcodegen}, self-improve efficiency through reinforcement learning~\cite{du2026afterburner}, and lift the optimization to project level~\cite{bai2025polo,ren2025peace}. Benchmarks frame this task~\cite{waghjale2024ecco,singhal2024nofuneval}, culminating in repository-level optimization that includes performance workloads and expert patches~\cite{he2025sweperf,shetty2026gso,ma2026swefficiency}. In contrast, we study the runtime performance of LLM-generated patches produced to resolve functional issues in repositories.

\paragraph{Efficiency-aware code generation.}
A parallel line of research injects efficiency into code generation, including self-optimization with execution profiles~\cite{huang2024effilearner}, efficiency-aware fine-tuning~\cite{huang2025efficoder}, LLM-as-critique prompting~\cite{zhu2026more}, preference alignment~\cite{zhang2025codedpo}, and reinforcement learning with performance rewards~\cite{feng2026towards}. The benchmarks pair programming problems with human-written solutions~\cite{huang2024effibench,qing2026effibench,du2024mercury}, commission expert-designed references~\cite{qiu2025efficient}, annotate asymptotic complexity~\cite{chambon2025bigo}, and rank solutions by instruction count~\cite{peng2025coffe}. They investigate the runtime performance of model-written code at the function or single-file level, rather than operating on real-world repositories.

\paragraph{Empirical studies of AI-generated code efficiency.}
Other work empirically studies the runtime cost of AI-produced code. Early work measures on HumanEval~\cite{chen2021evaluating}, MBPP~\cite{austin2021program}, and other tasks~\cite{niu2024evaluating,coignion2024performance,liu2024evaluating}. A series of studies document performance regressions in AI coding tools and trace their root causes~\cite{li2024assessing,li2026performance,erhabor2025measuring}, and categorize broader inefficiencies in LLM-generated code~\cite{abbassi2025taxonomy}. Some work compares mainstream tools and models across languages~\cite{guimaraes2025analyzing,li2026multi}. A recent study examines static characteristics of AI-generated code in real-world repositories~\cite{mao2026large}. In contrast, we analyze the runtime performance of LLM-generated patches for repository issues.

\section{Conclusion}
\label{sec:conclusion}

This study empirically analyzed the runtime performance of patches produced by LLM agents on repository-level issue-resolution benchmarks, using the benchmarks' built-in F2P tests as the workload. On the verified splits of SWE-bench and SWE-bench-Live, filtering for measurability left 339 instances, and 262 cases contained at least two genuinely different patches from four frontier models and the developer. We measured different solutions for each case on instruction counts and wall-clock times, and found that the built-in F2P tests carry genuine discrimination signal. Under Welch ANOVA, at most 42\% of cases show statistically distinguishable differences among their different correct patches. The distinguishable subsets the Welch ANOVA identifies are highly consistent with the full corpus for model comparisons. The measurement shows that issue resolution rate and runtime performance are two independent dimensions of model capability, with nearly reversed orderings. We call for future work to consider the performance of the LLM-generated patches in addition to issue resolution rate.

\begin{acks}

\end{acks}

\bibliographystyle{ACM-Reference-Format}
\bibliography{refs}

\appendix
\section*{Appendix}
\section{Decorator Implementation for Filtering}
\label{app:decorator}

The injection locates the F2P function definition in the test file by
matching its \texttt{def} statement and inserts \texttt{@measure\_all} before it. The decorator executes the function \texttt{WARMUP\_ITERS} times, emits a \texttt{PF\_TRACE\_START} marker, executes the measured pass, and emits \texttt{PF\_TRACE\_END}; this marker pair delimits the traced window.

\begin{lstlisting}[style=promptstyle]
WARMUP_ITERS = 10

def measure_all(func):
    caller_file = sys._getframe(1).f_code.co_filename

    if inspect.iscoroutinefunction(func):
        @functools.wraps(func)
        async def awrapper(*args, **kwargs):
            for _ in range(WARMUP_ITERS):
                await func(*args, **kwargs)
            os.write(2, b"PF_TRACE_START")
            ret = await func(*args, **kwargs)
            os.write(2, b"PF_TRACE_END")
            return ret
        try:
            awrapper.__code__ = awrapper.__code__.replace(co_filename=caller_file)
        except AttributeError:
            pass
        return awrapper

    @functools.wraps(func)
    def wrapper(*args, **kwargs):
        for _ in range(WARMUP_ITERS):
            func(*args, **kwargs)
        os.write(2, b"PF_TRACE_START")
        ret = func(*args, **kwargs)
        os.write(2, b"PF_TRACE_END")
        return ret
    try:
        wrapper.__code__ = wrapper.__code__.replace(co_filename=caller_file)
    except AttributeError:
        pass
    return wrapper
\end{lstlisting}

\section{Excluded Instances by In-process Repeatability}
\label{app:excluded}

\begin{table}[h]
  \centering
  \caption{Instances excluded by the in-process repeatability rule, by failure class.}
  \label{tab:excluded-classes}
  \footnotesize
  \begin{tabular}{l>{\raggedright\arraybackslash}p{0.72\linewidth}}
    \toprule
 Class & Case IDs \\
    \midrule
 Cannot execute (9) & astropy-8707, sphinx-9229, sphinx-13253,
 cryptography-12812, llama-factory-7505, fast-f1-699, pipenv-6380,
 llama-stack-1105, django-15525 \\
    \addlinespace
 Cannot isolate (12) & django-11532, privacyidea-4206, yt-dlp-11198,
 django-12209, cfn-lint-3653 / 3855 / 3856 / 3890, streamlink-6353,
 wemake-3117 / 3129, mypy-18966 \\
    \addlinespace
 Cannot repeat (66) & SWB (47): django-11066 / 11095 / 11138 / 11211 /
 11490 / 11551 / 11555 / 11734 / 11999 / 12143 / 12708 / 12774 /
 12858 / 13297 / 13315 / 13363 / 13401 / 13568 / 13809 / 13810 /
 14434 / 14752 / 14855 / 14999 / 15278 / 15315 / 15368 / 15561 /
 15695 / 15732 / 15814 / 16315 / 16333 / 16527 / 16631 / 16661,
 matplotlib-23299 / 25311 / 25332, pylint-4661, pytest-5809 / 7982 /
 10051, sphinx-8120 / 8269 / 8475 / 9711; SWBL (19): haystack-8941,
 pymdown-extensions-2651, flexget-4306, instructlab-2579 / 3118 /
 3135, dvc-10711, kedro-4540, flask-5637, patroni-3198, pdm-3255 /
 3314, smart\_open-833, privacyidea-4251, pylint-10036,
 python-control-1138, torchtune-1832, reflex-5077, linkding-1051 \\
    \bottomrule
  \end{tabular}
\end{table}

\noindent Per-class root-cause evidence for every case is catalogued
in the pipeline's hard-failure notes:
\begin{itemize}
  \item \textbf{Cannot execute (9).} External dependencies are unreachable
 at measurement time, a blocked package host (llama-factory-7505),
 a retired upstream API (fast-f1-699), internet-marked tests
 (pipenv-6380), a remote model endpoint (llama-stack-1105); or the
 dataset, the runner, or the platform is pathological, pytest no
 longer accepts the image's nose-style setup (xastropy-8707), the
 ground-truth patch does not produce the expected output (sphinx-9229), the
 runner hits a circular database dependency (django-15525), a
        \texttt{.pyi} stub fixture is not importable on Linux
 (sphinx-13253), and the ground-truth patch's binary content is truncated in
 the dataset (cryptography-12812). Several of them fail a
 file-level rerun just as well.
  \item \textbf{Cannot isolate (12).} The F2P identifier is
 corrupted, either truncated nodeids that reverse-derive to multiple F2P variants
 (cfn-lint-3653 / 3855 / 3856 / 3890, streamlink-6353,
 wemake-3117 / 3129) or a misrecorded \texttt{functools.partial}
 signature (django-12209); the F2P relies on state set by a
 preceding test in the same file, an object attribute
 (django-11532), database rows (privacyidea-4206), or an import
 side effect (yt-dlp-11198); or the test is data-driven and
 exposes no \texttt{def} function to instrument (mypy-18966).
  \item \textbf{Cannot repeat (66).} Each invocation mutates state that
 outlives it; by mechanism, database accumulation ($\sim$25),
 global or fixture state ($\sim$33), filesystem state ($\sim$6),
 and network hangs (2).
\end{itemize}

\section{Welch ANOVA Details}
\label{app:welch}

This section details the computation of the Welch statistic $F_W$ and
the Satterthwaite approximate degrees of freedom $\nu$ used in
\S\ref{sec:baseline-analysis}. Recall that $y_{i,g}=\ln x_{i,g}$ denotes
the log-transformed measurement of solution $i$ in replicate $g$, with
$k$ candidate solutions and $G$ replicates each.

For each solution $i$, its replicate mean and sample variance on the log
scale are
\begin{equation}
\bar{y}_i = \frac{1}{G}\sum_{g=1}^{G} y_{i,g},
\qquad
s_i^2 = \frac{1}{G-1}\sum_{g=1}^{G}\big(y_{i,g}-\bar{y}_i\big)^2,
\end{equation}
and the solution enters the analysis with the weight
$w_i = G/s_i^2$, which is inversely proportional to the variance of its
mean. Writing $W=\sum_{i=1}^{k} w_i$ for the total weight, the weighted
grand mean is
\begin{equation}
\tilde{y} = \frac{1}{W}\sum_{i=1}^{k} w_i\,\bar{y}_i.
\end{equation}
The Welch statistic sets the weighted between-solution spread against a
correction term $A$ that absorbs the variance heterogeneity,
\begin{equation}
F_W
= \frac{\dfrac{1}{k-1}\sum_{i=1}^{k} w_i\big(\bar{y}_i-\tilde{y}\big)^2}
 {1+\dfrac{2(k-2)}{k^2-1}\sum_{i=1}^{k}\dfrac{\big(1-w_i/W\big)^2}{G-1}},
\end{equation}
and the Satterthwaite approximate denominator degrees of freedom are
\begin{equation}
\nu
= \frac{k^2-1}
 {3\sum_{i=1}^{k}\dfrac{\big(1-w_i/W\big)^2}{G-1}}.
\end{equation}
Under the null hypothesis of equally fast solutions, $F_W$ approximately follows the $F$ distribution with $k-1$ and $\nu$ degrees of freedom, from which the $p$-value is obtained.

\section{RSD Distribution by Number of Distinct Solutions}
\label{app:rsd-by-k}
\begin{figure}[t]
  \centering
  \includegraphics[width=\linewidth]{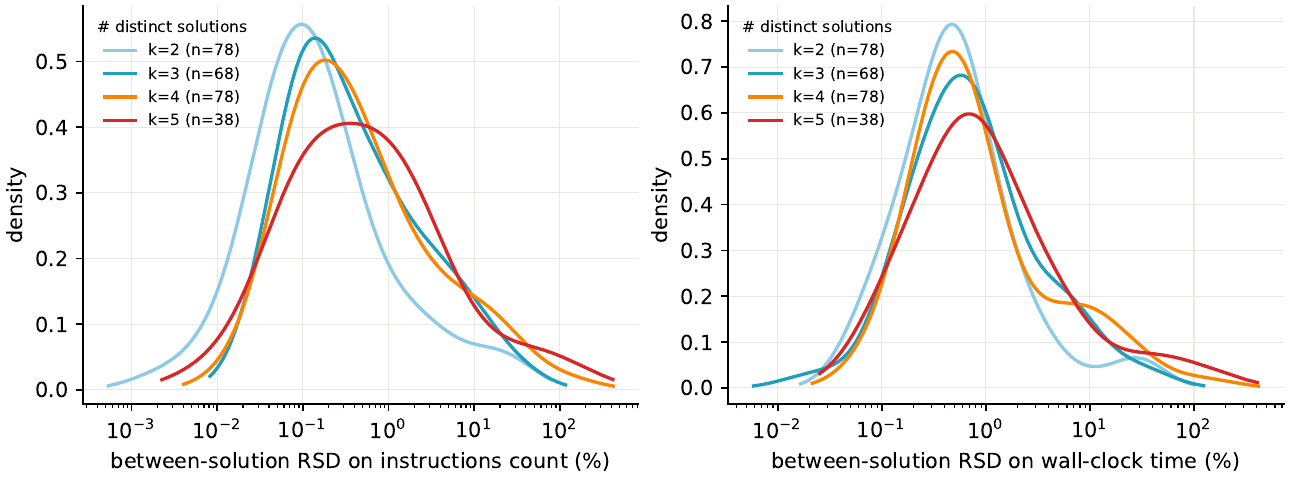}
  \caption{Density of the per-case between-solution RSD with F2P tests, grouped by the number of distinct solutions $k$, estimated with a kernel density smoother on the log scale of RSD. Left: instruction count. Right: wall-clock time.}
  \label{fig:rsd-by-k}
\end{figure}

Fig.~\ref{fig:rsd-by-k} shows the distribution of the per-case between-solution RSD, grouped by the number of distinct solutions $k$ admitted by the case. As $k$ grows, the distribution shifts right, its peak flattens, and its tail extends further; the same RSD value therefore sits at a different quantile in each group. A single threshold for all $k$ is therefore not feasible.

\section{Within-solution RSD and Case Study}
\label{app:within-rsd}
\begin{table}[h]
  \centering
  \caption{Within-solution RSD, median, mean, and max over cases.}
  \label{tab:within-rsd}
  \small
  \begin{tabular}{l ccc ccc}
    \toprule
    & \multicolumn{3}{c}{Instructions} & \multicolumn{3}{c}{Wall time} \\
    \cmidrule(lr){2-4} \cmidrule(lr){5-7}
 Dataset & Median & Mean & Max & Median & Mean & Max \\
    \midrule
 All (n=262) & 0.27\% & 0.49\% & 36.80\% & 0.82\% & 1.10\% & 36.41\% \\
 Dist. on instructions (n=110) & 0.25\% & 0.31\% & 1.48\% & 0.84\% & 0.93\% & 2.75\% \\
 Dist. on wall time (n=91) & 0.24\% & 0.32\% & 1.72\% & 0.80\% & 0.90\% & 2.75\% \\
    \bottomrule
  \end{tabular}
\end{table}

\noindent Table~\ref{tab:within-rsd} quantifies the within-solution run-to-run noise that the between-solution RSD ignores. We compute the within-solution RSD for each solution from its $G=5$ replicates, average the solutions' RSDs within each case, and summarize them as the median, mean, and max across cases. We find that the noise is stable but not negligible; the wall-time noise runs at 1--3\% and is two to three times the instruction-count noise. The full-set max, however, is much higher and comes from a single case, \emph{sympy\_\_sympy-20428}, whose within-solution RSD reaches 36.80\% on instruction count and 36.41\% on wall-time clock. The between-solution RSD of this case is about 10.8\%. It is exactly the isolated indistinguishable point at the right end of Fig.~\ref{fig:rsd-p-sorted} (above the 10\% level of the curves, a single point stays on the gray curve while colored segments cover its neighbors on both sides). It shows the highest between-solution RSD among all indistinguishable cases on both metrics.  

The case's two solutions come from Claude and the ground truth and share the same one-line core fix, which turns the zero test of the expression domain from a structural comparison into a symbolic one. In this case, we find clear bimodality in replicate costs, driven by per-process randomness. The test constructs a constant polynomial over a nested-radical expression and asserts it equal to the zero polynomial; under the patched zero test, the evaluation delegates to \emph{SymPy}'s assumption engine. Unable to decide the zero test structurally, the engine falls back to sign determination via minimal-polynomial computation and polynomial factorization over the integers. \textbf{This mechanism carries per-process randomness, so repeated executions fall into one of two modes whose costs differ by about 2$\times$.} Since each replicate is a fresh process, repeated executions of the same solution mix fast and slow runs, and the 36\% within-solution RSD is precisely this bimodal mixture. \textbf{The two solutions differ only by luck of the draw}; in our measurements, the model patch took the slow path in 1 of its 5 replicates, and the ground truth in 2 of 5, and this small difference alone produces a considerable between-solution RSD of 10.8\%. Considering the between-solution spread solely, RSD would read the case as a large genuine difference. The case is doubly unsuited for performance comparison: its measurement is unstable, and its two solutions hardly differ. The $p$-value weighs the between-solution spread against the within-solution noise, correctly declares the case indistinguishable, and removes it from the distinguishable subsets.

\end{document}